\documentclass[aps,superscriptaddress,amsmath,amssymb,showpacs,twocolumn,prl]{revtex4}
\usepackage{hyperref}
\usepackage{epsfig}
\usepackage{subfigure}
\usepackage{braket}
\usepackage{bm}
\usepackage{natbib}

\newcommand{\rem}[1]{}

\newcommand{\beq}{\begin{equation}}
\newcommand{\eeq}{\end{equation}}
\newcommand{\beqa}{\begin{eqnarray}}
\newcommand{\eeqa}{\end{eqnarray}}

\newcommand{\refe}[1]{\eqref{#1}}
\newcommand{\refE}[1]{Eq.~\eqref{#1}}

\newcommand{\eps}{\epsilon}
\newcommand{\Tr}{\mathrm{Tr}}

\begin{document}
\title{Andreev current induced by ferromagnetic resonance}

\author{Caroline Richard} 
\author{Manuel Houzet}
\author{Julia S. Meyer}
\affiliation{SPSMS, UMR-E 9001 CEA/UJF-Grenoble 1, INAC, Grenoble, F-38054, France}

\begin{abstract}
We study charge transport through a metallic dot coupled to a superconducting and a ferromagnetic lead with a precessing magnetization due to ferromagnetic resonance. Using the quasiclassical theory, we find that the magnetization precession induces a dc current in the subgap regime even in the absence of a bias voltage. This effect is due to the rectification of the ac spin currents at the interface with the ferromagnet; it exists in the absence of spin current in the superconductor. When the dot is strongly coupled to the superconductor, we find a strong enhancement in a wide range of parameters as compared to the induced current in the normal state.
\end{abstract}

\pacs{
74.45.+c,	%Proximity effects; Andreev reflection; SN and SNS junctions
75.76.+j,	%Spin transport effects
%76.50.+g,	%Ferromagnetic, antiferromagnetic, and ferrimagnetic resonances; spin-wave resonance
72.25.-b	%Spin polarized transport
}

\date{\today}

\maketitle

%\section{Introduction}

Spin-transfer torque allows one to manipulate the magnetization of a ferromagnetic (F) layer by means of a spin-polarized current  \cite{Slonczewski:1996,Berger:1996}. Random-access memories using this effect in order to induce magnetization reversal of the active elements are on their way to commercialization. The reverse effect, namely the generation of  a spin current in a normal metal (N) by means of a dynamically precessing ferromagnetic metal, has also been predicted \cite{Tserkovnyak:2002}. In the absence of direct spin probes, this effect may be measured by using a second ferromagnet as an analyzer that converts the spin current into a charge current. However, it was pointed out theoretically \cite{Wang:2006} and measured experimentally \cite{Costache:2006,Moriyama:2008} that a single F/N junction is enough to both generate and detect the spin current through the generation of a dc voltage at ferromagnetic resonance (FMR) in an open-circuit geometry. At the origin of this phenomenon is the  spin accumulation on the normal side of the junction -- due to the precession-induced spin current -- which is typically different from that on the ferromagnetic side. If transmissions for the majority and minority electron species through the junction are different, the difference in spin accumulation generates a net charge current which must be compensated by a difference in electrochemical potentials such that no charge accumulation occurs. Spin relaxation inhibits spin accumulation and, thus, suppresses the effect.

The aim of our work is to explore how this effect is modified in a ferromagnet/superconductor junction. The combination of ferromagnetic and superconducting (S) materials has been shown to lead to a variety of interesting spin phenomena \cite{Buzdin:2005,Bergeret:2005}. However, the study of the interplay between magnetization dynamics and superconductivity is a relatively new topic. Experimentally, a narrowing of the FMR width at the superconducting transition was observed in an F/S bilayer \cite{Bell:2008}. Theoretically, it was proposed that a dynamically precessing ferromagnet may generate a long-range proximity effect \cite{Houzet:2008}.  This effect would manifest itself in the enhancement of the critical current in a phase-biased ferromagnetic Josephson junction under FMR conditions. Related signatures in the tunneling density of states of the F layer have also been investigated \cite{Yokoyama:2009}. However, these works disregard interface effects and, therefore, do not take into account the possible generation of an FMR-induced dc voltage. Finally, let us note that the first experiments on voltage generation by FMR of Refs.~\cite{Costache:2006,Moriyama:2008} were performed with Al as the normal metal, which becomes superconducting at low temperatures. 

As the FMR-generated charge current in an F/N junction is typically associated with a spin current, one may wonder what happens in an F/S junction in the subgap regime, where transport is mediated by Andreev processes \cite{Blonder:1982}.  We show that the generation of charge current in the absence of a spin current in a conventional singlet superconductor is possible. In fact, the absence of spin currents in the superconductor may even lead to a strong enhancement of the induced charge current as compared to the normal state.

%\section{Model}

%%%%%%%%%%%%%%%%%%%%%%%%%%%%%%
%
%              FIGURE 1
%
%%%%%%%%%%%%%%%%%%%%%%%%%%%%%%
\begin{figure}[t]\centering
\includegraphics[width=.8\linewidth]{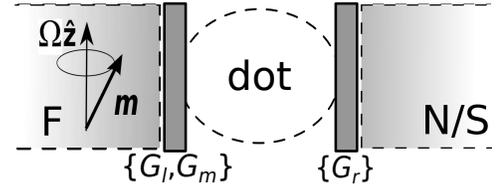}
\caption{
\label{fig1}
Setup of the junction. A metallic dot is coupled to a ferromagnetic lead with precessing magnetization $\bm{m}(t)$  on the left and to a normal or superconducting lead on the right. The left barrier is characterized by the conductances $G_l$ and $G_m$, defined above \refE{eq:BCleft}, whereas the right barrier is characterized by the conductance $G_r$.
}
\end{figure}
%%%%%%%%%%%%%%%%%%%%%%%%%%%%%%

As in the normal state, the two main ingredients necessary to generate the effect are spin-dependent transmissions through the junction and a spin accumulation region \footnote{The effect is absent in the theory of Ref. \cite{Skadsem:2011}, which did not include such a region.}. The simplest setup meeting these requirements is a metallic dot coupled through tunnel barriers to a ferromagnet and to a superconductor (see Fig.~\ref{fig1}).  Coulomb blockade effects are neglected, assuming that the conductances of the barriers largely exceed the conductance quantum.

The magnetization precession in the ferromagnetic lead is described by a time-dependent exchange field, $\bm{J}(t)=J\bm{m}(t)$ with
\beq
\bm{m}(t)=(\sin\theta\cos\Omega t,\sin\theta\sin\Omega t,\cos\theta),
\eeq
acting on the spin of the conduction electrons.
Here the precession frequency, $\Omega$, and the tilt angle, $\theta$, are both tunable with external dc and rf fields under standard FMR conditions \cite{Kittel:1948}. We will consider them as externally fixed parameters. 

The precession of the magnetization drives the system out of equilibrium and, thus, may generate a current. To describe the system, we use the quasiclassical Keldysh theory \cite{QuantumTransport}. In particular, the current through the junction can be expressed in terms of the quasiclassical Green function $\check g$ of the dot. Here $\check g$ is a matrix in Keldysh, Nambu, and spin space. In Keldysh space, it has a triangular structure with retarded ($\hat g^R$), advanced ($\hat g^A$), and Keldysh ($\hat g^K$) components. Furthermore, it satisfies the normalization condition $\check g^2=1$. 

The equations determining the Green functions are most conveniently written in the rotational frame for the magnetization precession, where the problem is stationary \cite{Houzet:2008}. 
Assuming that the conductance of the dot largely exceeds the conductances of the junctions with the leads, the equation determining $\check g$ 
may be cast in the form 
\beq
\label{eq:usadel}
-i\frac{2\pi G_Q}{\delta}
\left[\left(E+\frac\Omega 2\sigma_z \right)\tau_z, \check g 
\right]+\check I_l+ \check I_r=0.
\eeq 
Here $\sigma_i$ and $\tau _i$ are Pauli matrices in Nambu and spin space, respectively ($i=x,y,z$). Furthermore, $G_Q=e^2/\pi$ is the conductance quantum (in units where $\hbar=1$), and $\delta$ is the mean level spacing in the dot. The spin-dependent energy shift $\pm\Omega/2$ is a spin-resolved chemical potential induced  by the transformation from the laboratory to the rotational frame. The boundary conditions with the ferromagnetic ($l$ = left) and superconducting ($r$ = right) leads are represented by the matrix currents $\check I_{l/r}$ and depend on the Green functions $\check g_{l/r}$ describing the non-equilibrium state in the leads due to the magnetization precession.

Tunneling through an F/N interface is generally spin-dependent. The relevant processes can be characterized by the total conductance of the junction, $G_l$, and the difference between the conductances for the majority and minority electrons, $G_m$ \footnote{
In addition one may take into account the imaginary part of the spin-mixing conductance, $G_\phi$, which accounts for a spin-dependent phase shift upon reflection. In good F/N contacts with large Fermi velocity mismatch, $G_\phi$ is typically small and therefore neglected. It may, however, be large in tunnel junctions. As $G_\phi$ leads to spin relaxation, it would  suppress the effect studied in this work.}.
The matrix current at the tunnel interface between the dot and the ferromagnet then takes the form \cite{Huertas-Hernando:2002}
\begin{equation}
\label{eq:BCleft}
\check I_l=\frac{G_l}{2}[ \check g_l,\check g]
+\frac{G_m}{4}[ \{\bm{m}\cdot{\bm{\sigma}}\tau_z,\check g_l\},\check g]
,
\end{equation}
where $\bm{m}\equiv\bm{m}(0)$. Within the quasiclassical approximation, we assume $|G_m|\ll G_l$. Thus, $G_m$ can be treated perturbatively.

The Green function in the F lead, $\check g_l$, is determined by 
\beq
\label{eq:gleft}
\left[\left(E+\frac\Omega 2\sigma_z+J{\bm{m}}\cdot\bm{\sigma}\right)\tau_z
-\check \Sigma, \check g_l \right]=0,
\eeq 
where the self-energy $\check\Sigma=-i\Gamma \check g_N(E+(\Omega/2)\sigma_z)$ accounts for inelastic scattering in the relaxation time approximation. Here, $1/\Gamma$ is the inelastic scattering time and $\check g_N$ is the equilibrium Green function in a normal metal. Namely, $\hat g_N^{R(A)}(E)=\pm\tau_z$ and $\hat g_N^{K}(E)=2\tau_z f(E)$, where $f(E)=\tanh(E/2T)$ is related to the Fermi distribution at temperature $T$.
For a large exchange field, $J\gg \Omega,\Gamma$, the solution of \refE{eq:gleft} takes the form $\hat g_l^{R(A)}=\pm\tau_z$ and $\hat g_l^{K}=2\tau_z(f_++f_-\cos\theta\bm{m}\!\cdot\!{\bm{\sigma}})$, 
where $f_\pm(E)=[f(E+\Omega/2)\pm f(E-\Omega/2)]/2$.

The matrix current at the dot-superconductor tunnel interface is given as
\beq
\label{eq:BCright}
\check I_r=\frac{G_r}{2}[ \check g_r,\check g],
\eeq
where $G_r$ is the conductance of the junction. The Green function in the S lead reads 
$\check g_r=\check g_S(E+(\Omega/2)\sigma_z)$, where $\check g_S$ is the equilibrium Green function in a superconductor.
Namely, $\hat g^{R(A)}_S(E)=(-iE \tau_z+\Delta\tau_x)/\sqrt{\Delta^2-(E\pm i0^+)^2}$ 
and $\hat g^{K}_S(E)=[\hat g^{R}_S(E)-\hat g^{A}_S(E)]f(E)$,
where $\Delta$ is the superconducting order parameter (taken to be real). 

Now we have all the ingredients necessary to determine the Green function in the dot and subsequently the spin and charge currents at both interfaces.
The charge currents are given by
\beq
\label{eq:current}
I_{l/r}=\frac{1}{16e}\int dE\; \Tr[\tau_z \hat I_{l/r}^K].
\eeq
Current conservation ensures that $I\equiv I_l=-I_r$. 

The spin currents in the rotational frame are given by 
\beq
\label{eq:s-current}
\bm{I}_{l/r}=-\frac{1}{32e^2}\int dE \;\Tr[ \bm{\sigma} \hat I_{l/r}^K].
\eeq
In the laboratory frame, they decompose into a dc contribution along the precession axis, $I_{\alpha,z}$, and ac components in the perpendicular plane, $I_{\alpha,x/y}(t)=I_{\alpha,x/y}\cos\Omega t\mp I_{\alpha,y/x}\sin\Omega t$.
Contrarily to the charge current, the spin currents do not need to be conserved.
\refE{eq:usadel} yields
\beq
\bm{I}_{l}+\bm{I}_{r}-\frac{\Omega}{16\delta}
\int dE\;
\Tr
[
(\hat{\bf z}\times\bm{\sigma})\tau_z
\hat g^K
]=0.
\eeq
Thus, only the dc spin current along $\hat{\bf z}$ is conserved.

While our main interest are the FMR-induced currents in the subgap regime of an  F-dot-S junction, we first study the simpler case of an F-dot-N junction for comparison. For better readibility, in the following, we will normalize conductances by $G_\Sigma=G_l+G_r$ and energies by the Thouless energy $E_g=G_\Sigma\delta/(4\pi G_Q)$. In particular, we introduce the dimensionless conductances $\gamma_\alpha=G_\alpha/G_\Sigma$ ($\alpha=l,r,m$) as well as the dimensionless energies $\epsilon=E/E_g$ and $\omega=\Omega/(2E_g)$.

%\section{F-dot-N junction}

A normal lead is described by setting $\Delta=0$ in the above equations for $\check g_r$. In the absence of superconductivity, the retarded and advanced Green functions in the dot are trivial,
$\hat g^{R(A)}=\pm\tau_z$. The Keldysh component is obtained with the help of Eqs.~\refe{eq:usadel}, \refe{eq:BCleft}, and \refe{eq:BCright}. 
There is a remarkable relation between the spin current at the left contact with the F lead
and the charge current to lowest order in $\gamma_m$, 
\beq
\label{eq:Ics}
I=\frac{2e\gamma_m\gamma_r}{\gamma_l}\bm{m}(t)\cdot\bm{I}_l(t)=\frac{2e\gamma_m\gamma_r}{\gamma_l}(I_{l,x}\sin\theta+I_{l,z}\cos\theta),
\eeq
namely the charge current is proportional to the projection of the spin current onto the instantaneous magnetization axis of the barrier due to the spin-dependent conductance $G_m$.
That is, the charge current originates from two effects: (i) the rectification of the ac in-plane spin current pumped from the ferromagnet, and (ii) the conversion of the dc spin current along the $\hat{\bf z}$-axis into a charge current.
It turns out that the two effects have opposite sign, and that the former dominates over the latter. Namely,
we find
\begin{subequations}
\begin{eqnarray}
I_{l,x}&=&\frac{G_lE_g}{2e^2}\frac{\omega(\gamma_r+\omega^2)}{1+\omega^2}\sin\theta\cos\theta,\\
I_{l,z}&=&-\frac{G_lE_g}{2e^2}\omega\gamma_r\sin^2\theta.
\end{eqnarray}
\end{subequations}
Note that, in the limit $\gamma_r\ll \gamma_l$, the spin current along the $\hat{\bf z}$-axis is negligible. By contrast,  in the limit $\gamma_l\ll\gamma_r$, the two components are of comparable magnitude and almost complete cancellation between the competing effects takes place.

The charge current reads
\beq
\label{eq:IFdotN1}
I=\frac{G_m E_g}{e}
\frac{\gamma_r\gamma_l\omega^3\sin^2\theta\cos\theta }
{1+\omega^2}.
\eeq 
At large precession frequency, $\omega\gg 1$, the current scales linearly with  frequency,
$I\simeq (G_lG_rG_m/2eG_\Sigma^2)\Omega \sin^2\theta\cos\theta$. In particular, in an open-circuit geometry, this would correspond to an FMR-induced dc voltage
$eV=(G_m/2G_\Sigma)\Omega \sin^2\theta\cos\theta$ in accordance with Refs. \cite{Wang:2006,Costache:2006,Moriyama:2008}.
At $\omega\ll 1$, spin-relaxation mechanisms induced by the tunnel coupling of the dot to the leads tend to suppress the effect.

%\section{F-dot-S junction}

We now turn to the F-dot-S junction. In the subgap regime, the spin current at the interface with the superconductor vanishes, $\bm{I}_r=0$. Thus, $I_{z,l}=0$. However, an ac spin current is present at the interface with the ferromagnet. Then, the Andreev charge current originates entirely from the rectification of this ac spin current. 

Restricting ourselves to energy scales much smaller than $\Delta$, the Green function in the superconducting lead takes the simple form $\hat g_r^{R(A)}=\tau_x$ and $\hat g_r^K=0$.
Taking $\gamma_m$ as a small parameter, we search for a perturbative solution of equation \refe{eq:usadel} in the form $\check g =\check g_0+\gamma_m\check g_1+\dots$ 

Due to the proximity effect, now the retarded and advanced Green functions of the dot are modified as well. To zeroth order in $\gamma_m$, an explicit solution is given by
\beq
\label{eq:gs0}
\hat g_0^{R(A)}=
\frac{\gamma_r\tau_x+[-i(\eps+\omega\sigma_z)\pm \gamma_l]\tau_z}
{\sqrt{\gamma_r^2-(\eps+\omega\sigma_z\pm i\gamma_l)^2}}.
\eeq
Here, $\gamma_r$ is the effective minigap due to the coupling with the S lead \cite{Beenakker:2005}, $\omega$ an effective exchange field, and $\gamma_l$ yields a broadening of the energy levels due to the coupling with the F lead.
The Keldysh Green function can be cast in the form
$\hat g_0^K=\hat g^R\hat\varphi-\hat\varphi\hat g^A$ with 
\beq
\hat\varphi
=
f_++f_-\cos\theta\left[
\frac{\gamma_l\sin\theta}{\omega^2+\gamma_l^2}
(\gamma_l\sigma_x\!-\!\omega\sigma_y)
+\cos\theta\sigma_z\right]\!\!.
\eeq
The function $\hat\varphi$ can be interpreted as a matrix-distribution function. Note that it does not depend on $\gamma_r$ as subgap electrons only thermalize with the F lead.

\begin{figure}[t]\centering
\includegraphics[width=0.85\linewidth]{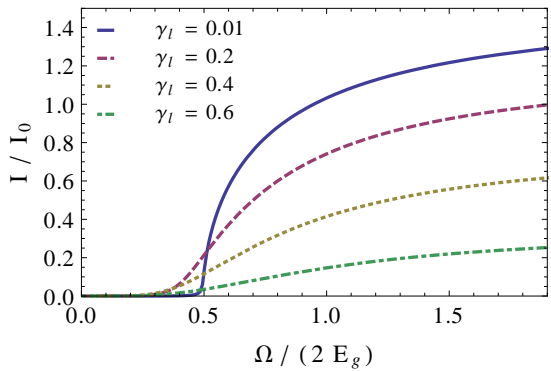}
\caption{
\label{fig2}
Andreev current induced by ferromagnetic resonance as a function of precession frequency for different values of $\gamma_l=G_l/(G_r+G_l)$. 
Here $I_0=(G_mE_g/e)\sin^2\theta\cos\theta$.
}
\end{figure}

To first order in $\gamma_m$, a solution which satisfies the normalization condition, $\check g^2=1$, is obtained in the form
$\check g_1=\check g_0 \check X -\check X \check g_0$. For the advanced and retarded components, one finds 
$\hat X^{R(A)}=\mp (\sin\theta/2 \omega)[i\gamma_r/(\epsilon\pm i\gamma_l) \tau_x
+\tau_z]\sigma_y$.
%\beq
%\hat X^{R(A)}
%=
%\mp \frac{\sin\theta}{2 \omega}
%\left(
%i\frac {\gamma_r}{\epsilon\pm i\gamma_l} \tau_x
%+\tau_z
%\right)\sigma_y.
%\eeq
The Keldysh component can be decomposed as $\hat X^K=X^K_x\tau_x+X^K_z\tau_z$,
where $X^K_x$ and $X^K_z$ solve the coupled equations
\begin{subequations}
\label{eq:X}
\beqa
&&2 \gamma_l X^K_z -i\omega[\sigma_z,X^K_z]\\&=&2 \sin\theta[ \cos\theta \sin\theta f_-
 + f_+(\sigma_x-\frac { \gamma_l}{ \omega}\sigma_y)],\nonumber
\\
&&2\epsilon X^K_x-2i \gamma_r X^K_z+\omega\{\sigma_z,X^K_x\}\\&=&
 2i\frac {\gamma_l\gamma_r\sin\theta}{ \omega(\gamma_l^2+ \epsilon^2)}
 [\gamma_l f_+\sigma_y-\epsilon f_- \cos\theta(\cos\theta\sigma_x-\sin\theta \sigma_z) ].\nonumber
 \eeqa
\end{subequations}
Evaluating the current at the right interface,  \refE{eq:current} yields $I=-i\gamma_m G_r/(16e)\int dE\; \Tr[\tau_y \hat g_1^K]$. Inserting the solution for $\hat g_1^K$ and using the property $\hat g_0^R(-\epsilon)=-\sigma_x\tau_z\hat g_0^A(\epsilon)\sigma_x\tau_z$, we obtain the current
\beqa
\label{eq:res}
I&=& \frac{1}{2}
I_0\frac{\gamma_r^2\omega }{\gamma_l^2+\omega^2}
\int d\epsilon\;
\frac{\epsilon f_-}{(\gamma_l^2+\epsilon^2)(\epsilon+\omega)}
\\
&&\qquad\qquad\qquad \times
\sum_\pm \frac{-\gamma_l(\epsilon+\omega)\pm i(\gamma_l^2-\epsilon\omega)}
{\sqrt{\gamma_r^2-(\epsilon+\omega \pm i \gamma_l)^2}},\nonumber
\eeqa
where $I_0= (G_mE_g/e)\sin^2\theta\cos\theta$. The current as a function of frequency for different values of $\gamma_l=1-\gamma_r$ is shown in Fig.~\ref{fig2}. Simple analytic expressions  can be found in different asymptotic regimes. 

%%%%%%%%%%%%%%%%%%%%%%%%%%%%%%
%
%              FIGURE 2
%
%%%%%%%%%%%%%%%%%%%%%%%%%%%%%%
%%%%%%%%%%%%%%%%%%%%%%%%%%%%%%

In particular, at temperature $T=0$  and low frequency $|\omega|\ll\gamma$, where $\gamma=(\gamma_l^2+\gamma_r^2)^{1/2}$, the FMR-induced current is given as
\beq
\label{eq:FdotS:lowfreq}
I\simeq
\frac{10}{3}
I_0
\frac{\gamma_r^2\gamma_l}{\gamma^7}\omega^5.
\eeq
The large power $\omega^5$ indicates the strong suppression of the effect.

At large frequencies $|\omega|\gg\gamma$, the current saturates. The frequency-independent value is given by
\beq
\label{eq:S-omegalarge}
I\simeq
\frac{\pi}{2}I_0\,\mathrm{sign}(\omega)
\times
\left\{\begin{array}{lll}
\gamma_r^2, & \,& \gamma_r\ll \gamma_l, \\
1, & \,& \gamma_l\ll \gamma_r.
\end{array}
\right.
\eeq
The saturation can be understood as Andreev processes become inefficient at energies larger than the minigap \cite{Volkov:1993}.

Depending whether the dot is more strongly coupled to the ferromagnet or to the superconductor, the crossover between these asymptotic regimes is different. If the dot is weakly coupled to the superconductor, $\gamma_r\ll \gamma_l$, a smooth crossover happens at $\omega\sim1\gg\gamma_r$ with a typical current
$I/I_0\sim \gamma_r^2$. By contrast, if the dot is weakly coupled to the ferromagnet, $\gamma_l\ll \gamma_r$, the crossover in the region $\omega\sim 1/2$ is described by
\beq
\label{eq:crossover}
I\simeq
I_0
\times\!
\left\{
\begin{array}{lll}
\gamma_l/(2\sqrt{-\delta\omega}),
&
\,
&\!
-1\ll \delta\omega\ll-\gamma_l,
\\
2\sqrt{\delta\omega},
&
\,
&\!
\gamma_l\ll \delta\omega\ll1,
\end{array}
\right.
\eeq
where $\delta\omega\equiv\omega-1/2$, with a typical current $I/I_0\sim \sqrt{\gamma_l}$ at $\omega=1/2$.

\begin{figure}[t]\centering
\includegraphics[width=0.85\linewidth]{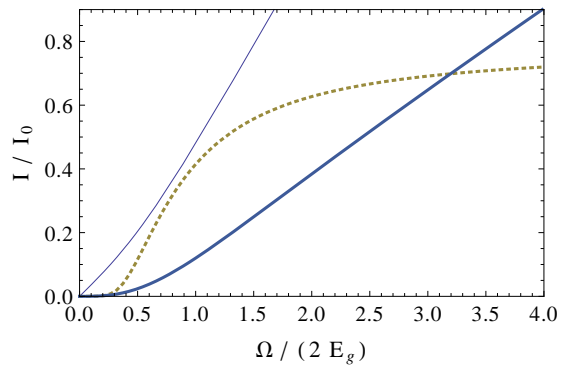}
\caption{
\label{fig3}
Induced current in the superconducting (dotted line) and normal state for $\gamma_l=0.4$. 
The thin line shows the contribution to the normal state current due to rectification only.}
\end{figure}

While in the asymptotic regimes of $\omega$ very small or very large, the current is suppressed as compared to the normal state, there is in fact a wide intermediate regime where it may be strongly enhanced. Comparing Eqs.~\eqref{eq:IFdotN1} and \eqref{eq:S-omegalarge}, one notices that, if the dot is strongly coupled to the superconductor, in the regime $1/2<\omega<\gamma_l^{-1}$  the induced current in the superconducting state exceeds the induced current in the normal state. This effect may be understood due to the absence of a dc spin current along along the $\hat{\bf z}$-axis which leads to a strong suppression of the effect in the normal state. 
Fig.~\ref{fig3} shows the current in the superconducting and normal state as well the contribution due to rectification only in the normal state. The ratio between the current in the superconducting state and the latter contribution in the normal state reflects the ratio between Andreev and normal state conductances in an N-dot-S junction.

%%%%%%%%%%%%%%%%%%%%%%%%%%%%%%
%
%              FIGURE 3
%
%%%%%%%%%%%%%%%%%%%%%%%%%%%%%%
%%%%%%%%%%%%%%%%%%%%%%%%%%%%%%

%\section{Inhomogeneous magnetization}

So far we assumed that the magnetization in the ferromagnet is uniform. However, boundary effects may lead to a suppression of the magnetization in the vicinity of the F/N interface. This  would result in a different resonance frequency at the barrier than in the ferromagnetic reservoir and, consequently, in a tilt angle $\theta_B\neq\theta$ at FMR. The effect can be accounted for by replacing $\bm{m}$ with $\bm{m}_B=(\sin\theta_B,0,\cos\theta_B)$ in \refE{eq:BCleft}. In particular, at $\theta_B=0$, the spin dependent conductance $G_m$ refers to the constant axis $\hat{\bf  z}$.  

In the normal case, the relation between spin and charge currents, \refE{eq:Ics},  now reads $I=(2e\gamma_m\gamma_r/\gamma_l)\bm{m}_B(t)\cdot\bm{I}_l(t)$. While at $\theta_B=\theta$ the rectification of the in-plane ac spin currents always dominates over the conversion of the dc spin current along $\hat{\bf  z}$ into a charge current, this effect is completely suppressed at $\theta_B=0$. As a consequence, at $\theta_B=0$, the charge current, 
$I=-({G_m E_g}/{e})\omega \gamma_r^2\sin^2\theta$,
has the opposite sign compared to \refE{eq:IFdotN1}.
In general, both effects are important. 
The sign reversal occurs at
$\tan\theta_B=[1-\gamma_l\omega^2/(\gamma_r+\omega^2)]\tan\theta$.

In the superconducting case, the dc spin current along $\hat{\bf  z}$ is always zero, and the charge current is due entirely to the rectification of the in-plane ac spin currents. As a consequence, we find that the charge current vanishes at $\theta_B=0$. The general result is obtained from \refE{eq:res} by replacing $I_0$ with $I_0^B=(G_mE_g/e)\sin\theta_B\sin\theta\cos\theta$. 

In summary, we demonstrate that a subgap charge current in an F/S junction may be induced by ferromagnetic resonance. The effect is due to the rectification of ac spin currents generated by the precessing magnetization in the ferromagnet. In the normal case, a competing effect of conversion of  a dc spin current into a charge current exists. This effect is absent in an F/S junction as the superconductor cannot carry a subgap spin current. As a consequence, the induced current in the superconducting state may be strongly enhanced as compared to the normal state. Interesting non-equilibrium phenomena should be expected in ferromagnetic Josephson junctions under ferromagnetic resonance conditions.

\acknowledgements
We acknowledge funding through an ANR grant (ANR-11-JS04-003-01) and an EU-FP7 Marie Curie IRG.

\bibliographystyle{apsrev4-1}
\bibliography{biblio}
\end{document}